\documentclass[letter,11pt]{article}
\pdfoutput=1 

\usepackage{jcappub} 

\usepackage[normalem]{ulem}

\usepackage[utf8]{inputenc}
\usepackage[T1]{fontenc} 
\usepackage{dsfont}
\usepackage{amsmath,amssymb,calc}
\usepackage{color}
\usepackage{amssymb}
\usepackage{graphicx, epsfig, bm}
\usepackage{hyperref}
\usepackage{soul}
\usepackage{multicol}
\usepackage{changepage}
\usepackage{subfig}

\usepackage{color}
\usepackage{ifthen}

\def\bdm{\begin{displaymath}}
\def\edm{\end{displaymath}}

\def\barray{\begin{array}}
\def\earray{\end{array}}
\def\be{\begin{equation}}
\def\ee{\end{equation}}
\def\ben{\begin{equation} \nonumber}
\def\een{\end{equation}}
\def\ban{\begin{eqnarray*}}
\def\ean{\end{eqnarray*}}
\def\ba{\begin{eqnarray}}
\def\ea{\end{eqnarray}}
\def\eal{\end{align}}
\def\bal{\begin{align}}

\def\({\left(}
\def\){\right)}
\def\[{\left[}
\def\]{\right]}

\def\by{{\bf{y}}}
\def\ba{{a\left(\by\right)}}

\def\bk{{\bf k}}

\def\bx{{\bf x}}

\definecolor{gold}{rgb}{1.0, 0.84, 0.0}
\definecolor{maroon}{rgb}{.25,0,0}
\definecolor{darkorange}{rgb}{1.0, 0.55, 0.0}
\definecolor{corn}{rgb}{0.98, 0.93, 0.36}
\definecolor{bronze}{rgb}{0.8, 0.5, 0.2}
\definecolor{darkgreen}{cmyk}{0.85,0.2,1.00,0.2}

\begin{document}

\title{Spectra of fermions produced by a time-dependent axion in the radiation- and matter-dominated Universe}

\author{Aditya Kulkarni}
\affiliation{Amherst Center for Fundamental Interactions, Department of Physics, University of Massachusetts, Amherst, MA 01003, U.S.A.}
\author{and Lorenzo Sorbo}




\abstract{Axion-like degrees of freedom generally interact with fermions through a shift symmetric coupling.  As a consequence, a time-dependent axion will lead to the generation of fermions by amplifying their vacuum fluctuations. We provide the formulae that allow one to determine the spectra of produced fermions in a generic Friedmann--Lema\^\i tre--Robertson--Walker Universe with flat spatial slices. Then we derive simple approximate formulae for the spectra of the produced fermions, as a function of the model parameters, in the specific cases of a radiation- and a matter-dominated Universe, in the regime in which the backreaction of the produced fermions on the axionic background can be neglected.}

\maketitle


\section{Introduction}%

Axion-like particles are well-motivated candidate degrees of freedom for the description of the physics beyond the Standard Model. Thanks also to the radiative stability of their potential, which is guaranteed by a softly broken shift symmetry, axions can play a significant role in cosmology: axion-like fields are excellent inflaton candidates~\cite{Freese:1990rb,Silverstein:2008sg,McAllister:2008hb,Kaloper:2008fb,Kaloper:2011jz}; the QCD axion (or more in general an axion-like particle with a sufficiently large mass) might constitute Dark Matter~\cite{Preskill:1982cy,Abbott:1982af,Dine:1982ah}; last but not least, an axion-like degree of freedom with a mass of the order of $10^{-33}$~eV is a natural candidate for a dynamical explanation of the current accelerated expansion of the Universe~
\cite{Hill:1988vm,Frieman:1991tu,Frieman:1995pm}.  An arrangement of various axions might even play all of these roles in the ``axiverse'' scenario~\cite{Arvanitaki:2009fg,Bachlechner:2019vcb}.

The interactions of cosmological axion-like degrees of freedom with other forms of matter have been studied for several decades. In this work, we will be concerned with the fact that a rolling homogeneous axion can provide a time-dependent background for other fields, thus leading to amplification of their vacuum fluctuations. In this respect, the phenomenology of a homogeneous axion coupled to gauge fields and rolling either during inflation (see e.g.~\cite{Pajer:2013fsa} for a review) or in the post-inflationary Universe (see e.g.~\cite{Agrawal:2017eqm,Kitajima:2017peg,DallAgata:2019yrr,Berghaus:2019cls}) has been studied in the detail.

More recently, several works, including~\cite{Adshead:2015jza,Adshead:2015kza,Adshead:2018oaa,Adshead:2019aac,Domcke:2021fee,Roberts:2021plm,Domcke:2019qmm,Domcke:2018eki}, have focused on the fact that a homogeneous rolling axion can lead to matter production through a shift-symmetric coupling to fermions. This coupling has received comparatively less interest than the coupling to gauge fields, which can be attributed to the fact that, due to Pauli blocking, fermions do not achieve the exponentially large occupation numbers obtained by vectors~\cite{Anber:2006xt}. In the case of fermion production, the limitations from Pauli blocking can however be overcome by filling the Fermi sphere up to large momenta, leading to a phenomenology that is dramatically different, as one could expect, from that of vectors. In particular, the authors of~\cite{Adshead:2015jza,Adshead:2015kza,Adshead:2018oaa,Adshead:2019aac,Roberts:2021plm} have studied the production of fermions during inflation. The more recent papers~\cite{Domcke:2021fee,Domcke:2019qmm,Domcke:2018eki}, focus mostly  on the effects of a rolling axion in the presence of a uniform electromagnetic field. Nevertheless, they also contain analyses of fermion production by a rolling axion in the absence of gauge fields, on a Minkowskian background.

In the present work, we will consider fermion production by a time-dependent homogeneous axion in a general flat Friedmann-Lema\^\i tre-Robertson-Walker (FLRW) cosmology. In particular, after deriving the formulae controlling the evolution of the Bogolyubov coefficients, we will specialize to the case of axion-like fields oscillating in a radiation- and slowly rolling in a matter-dominated background. We use these equations to provide simple general expressions for the spectra of the fermions generated in this system. These spectra are obtained by integrating analytically the equations for the Bogolyubov coefficients in the regime in which the occupation number is much smaller than unity. Due to Pauli blocking, when this condition is not realized, we know that the occupation number must be of the order of the unity. We numerically check the validity of our analytical results. Throughout the paper we ignore the backreaction of the produced fermions on the axion. At the end of the paper, we list a few scenarios where these formulae (or their appropriate extensions) could be applied.

Our work is organized as follows. In Section~\ref{sec:equations} we present our model and derive the expressions of the Bogolyubov coefficients and the equations that govern them in a general flat FLRW space. In Section~\ref{sec:solutions} we solve numerically those equations for radiation-dominated and matter-dominated backgrounds and present simple formulae for the scaling of the total number of produced fermions as a function of the parameters of the model. In Section~\ref{sec:summary} we discuss some possible applications of the results presented in Section~\ref{sec:solutions} and summarize our work. Estimating the fermion spectra requires analytical estimates of integrals that have to be performed in different regions of the parameter space. We review an example of such calculations in the Appendix.

\section{Deriving the equation for Bogolyubov coefficients}
\label{sec:equations}%

In this work we consider an axion-like field $\phi$ with axion constant $f$ interacting with a fermion $\psi$ through a dimension-five derivative coupling as well as a mass term with an exponential dependence on $\gamma_5\,\phi$. These two interaction terms come as a pair, as one can be converted into the other by a redefinition of the fermion field (see below, right after eq.~(\ref{eq:Dirac1})). We work in a flat FLRW Universe, with the scale factor denoted by $a$. The relevant part of our action is
\begin{equation}
{\cal S}=\int d^4x\,a^4\left[\Bar{\psi}\left(i\frac{\gamma^\mu}{a}\partial_\mu + \frac{3}{2}i\frac{a'}{a^2}\gamma^0-m\,e^{2i\gamma_5\,c_m\phi/f}+c_5\frac{\gamma^\mu}{a}\gamma_5\frac{\partial_\mu\phi}{f}\right)\psi+\frac{1}{2}\partial_{\mu}\phi\partial^{\mu}\phi-V(\phi)\right]\,,
\end{equation}
where we use conformal time, which we denote by $\tau$, where $\gamma^\mu$ are gamma matrices in Minkowski spacetime in the chiral representation, 
\begin{align}
    \gamma^0=\left(
    \begin{array}{cc}
    0  & 1_2 \\
    1_2  &  0
    \end{array}\right)\,,\qquad
    \gamma^i=\left(
    \begin{array}{cc}
    0  & \sigma_i \\
    -\sigma_i  &  0
    \end{array}\right)\,,\qquad \gamma_5=i\gamma^0\,\gamma^1\,\gamma^2\,\gamma^3=\left(
    \begin{array}{cc}
    -1_2  & 0 \\
    0  &  1_2
    \end{array}\right)\,,
\end{align}
(with $1_2$ denoting the $2\times 2$ identity matrix), and where $c_5$ and $c_m$ are dimensionless coupling constants of the dimensionless axion $\phi/f$. Also, a prime denotes derivative with respect to $\tau$. We simplify our notation by performing the substitutions $\frac{c_5}{f}\phi\rightarrow\theta_5$ and $\frac{c_m}{f}\phi\rightarrow\theta_m$.

The fermion equation of motion reads
\begin{equation}\label{eq:Dirac1}
		\left(i\gamma^\mu\partial_{\mu}+\frac{3}{2}i\frac{a'}{a}\gamma^0-m\,a\,e^{2i\gamma_5\theta_m}+{\gamma^\mu}\gamma_5\,\partial_{\mu}\theta_5\right)\psi = 0\,.
\end{equation}

Further substituting $\psi \rightarrow e^{-i\gamma_5\theta_5}\psi$ , $\psi\rightarrow a^{-3/2}\psi$ and defining $\theta_5 + \theta_m \equiv \theta$, the equation of motion simplifies drastically to
\begin{equation}
		\left(i\gamma^\mu \partial_{\mu}-m\,a\,e^{2i\gamma_5\,\theta}\right)\psi = 0\,.
\end{equation}

We wish to solve this equation for $\theta=\theta(\tau)$. To do so, we decompose the fermionic field function in Fourier modes as
\begin{equation}\label{eq:spinordeco1}
		\psi(\textbf{x},\,\tau)\equiv\int \frac{d^3\bk}{(2\pi)^{3/2}}e^{i\bk\cdot\bx}\psi_\bk(\tau)=\int \frac{d^3\bk}{(2\pi)^{3/2}}e^{i\bk\cdot\bx}\sum_{r=\pm}\left[U_r(\bk,\,\tau)\,\hat{a}_r(\textbf{k})+V_r(-\textbf{k},\,\tau)\,\hat{b}_r^{\dagger}(-\textbf{k})\right]\,,
	\end{equation}
where
\begin{equation}\label{eq:spinordeco2}
	U_r(\textbf{k},\,\tau)=\frac{1}{\sqrt{2}}
	\begin{pmatrix}
		\chi_r(\hat\bk)\,u_r(k,\,\tau)\\
		r\,\chi_r(\hat\bk)\,v_r(k,\,\tau)
	\end{pmatrix}, \qquad 
	V_r(-\textbf{k},t)=\frac{1}{\sqrt{2}}
	\begin{pmatrix}
		\chi_r(\hat\bk)\,w_r(k,\,\tau)\\
		r\,\chi_r(\hat\bk)\,y_r(k,\,\tau)
	\end{pmatrix}\,,
\end{equation}
and with the spinors $\chi_r(\hat\bk)$ defined as
\begin{equation}
	\chi_r(\hat\bk)=\frac{(1+r\,\pmb{\sigma}\cdot\hat\bk)}{\sqrt{2(1+\hat{\bk}_3)}}\Bar{\chi}_r,\qquad\bar{\chi}_{+}=\begin{pmatrix}1\\0\end{pmatrix},\,\qquad\bar{\chi}_{-}=\begin{pmatrix}0\\1\end{pmatrix}\,,
\end{equation}
that are orthonormal helicity eigenstates:
\begin{align}
    \pmb{\sigma}\cdot \hat\bk\,\chi_r(\hat\bk)=r\,\chi_r(\hat\bk)\,,\qquad\qquad \chi^{\dagger}_r(\hat\bk)\chi_s(\hat\bk)=\delta_{rs}\,.
\end{align}

Using these definitions and properties, we get the equations of motion for the functions $u_r$ and $v_r$  as
\begin{align}\label{eq:eom_uv}
		&i\,u'_r+r\,k\,u_r-m\,a\,r\,e^{2i\theta}\,v_r =0\,,\nonumber\\
		&i\,v'_r-r\,k\,v_r-m\,a\,r\,e^{-2i\theta}\,u_r=0\,,
\end{align}
where the quantity $|u_r|^2+|v_r|^2$ is a constant of motion that we normalize to $|u_r|^2+|v_r|^2=2$.

The expression for $w_r$ and $y_r$ can be obtained by observing that, as a consequence of the invariance of the system under charge conjugation,  $V_r(-\textbf{k})$ satisfies the same equation as $i\gamma^0\gamma^2\Bar{U}_r(\textbf{k})^T$. Then, using the properties $i\sigma_2\,\chi_r^*({\bf k})=-r\,\chi_{-r}({\bf k})=-e^{-ir\phi_{\bf k}}\chi_r({\bf k})$, with  $e^{i\phi_\bk}=\frac{k_1+ik_2}{\sqrt{k_1^2+k_2^2}}$, we can identify $w_r=-r\,v_r^*$ and $y_r=r\,u_r^*$, up to the irrelevant constant phase $e^{-ir\phi_{\bf k}}$.

The Bogolyubov coefficients can be found by diagonalizing the Hamiltonian which, after rotating away\footnote{If we did not rotate away $\theta_5$, the fermionic Hamiltonian would contain an extra term $i\int d^3\bk\,\psi^{\dagger}_\bk\,\gamma_5\theta'_5\psi_\bk$, which originates from the fact that $\partial_\mu\theta_5\supset \phi'$, which modifies the expression of the momentum conjugate to $\phi$. Such a term is essential to keep the theory, including the evolution of the Bogolyubov coefficients, invariant under the symmetry $\psi\mapsto e^{-i\gamma_5\alpha}\psi$, $\theta_5\mapsto \theta_5-\alpha$, $\theta_m\mapsto\theta_m+\alpha$.} $\theta_5$, reads
\begin{equation}
	\hat{H} = i\int d^3\bk\,\psi^{\dagger}_\bk\,\psi_\bk'\,,
\end{equation}
and that, after substituting the spinor decomposition~(\ref{eq:spinordeco1}) with~(\ref{eq:spinordeco2}), takes the form
\begin{equation}\label{eq:ham_non_diag}
	\hat{H} = \int d^3\bk 
	\left(\hat{a}^{\dagger}_r(\bk),\, \hat{b}_r(-\bk)\right)
	\begin{pmatrix}
		{\cal A}_r(k,\,\tau) & {\cal B}^*_r(k,\,\tau) \\ 
		{\cal B}_r(k,\,\tau) & - {\cal A}_r(k,\,\tau)
	\end{pmatrix}
	\begin{pmatrix}
		\hat{a}_r(\bk) \\ 
		\hat{b}^{\dagger}_r(-\bk)
	\end{pmatrix}\,,
\end{equation}
where
\begin{align}
	\begin{split}
	{\cal A}_r(k,\,\tau) &= \frac{r}{2}\left[-k\,|u_r|^2 + k\,|v_r|^2+m\,a\,e^{2i\theta}\,u^*_r\,v_r + m\,a\,e^{-2i\theta}\,v^*_r\,u_r\right]\, \\
	{\cal B}_r(k,\,\tau) &= \frac{1}{2}\left[2\,k\,u_r\,v_r+m\,a\,e^{-2i\theta}\,u_r^2 - m\,a\,e^{2i\theta}\,v_r^2\right]\,.
	\end{split}
\end{align}

The fact that the Hamiltonian is not diagonal (i.e., that it includes terms proportional to, e.g., $\hat{a}_r^\dagger(\bk)\,\hat{b}_r^\dagger(-\bk)$) implies that in general the operators $\hat{a}^\dagger_r(\bk)$ and $\hat{b}^\dagger_r(\bk)$ cannot be interpreted as creation operators of physical energy eigenstates. Physical creation/annihilation operators are then found by diagonalizing the Hamiltonian. 

A direct calculation shows that the eigenvalues of the $2\times 2$ matrix appearing in eq.~(\ref{eq:ham_non_diag}) are $\pm\omega_k(\tau)$, $\omega_k(\tau)\equiv \sqrt{k^2+m^2\,a(\tau)^2}$, so that the diagonalization will be realized by finding two functions $\alpha_r(k,\,\tau)$ and $\beta_r(k,\,\tau)$  for which $|\alpha_r|^2 + |\beta_r|^2 = 1$ and
\begin{eqnarray}
\left(\begin{array}{cc}
{\cal A}_r & {\cal B}_r^* \\ 
{\cal B}_r & -{\cal A}_r 
\end{array}\right)=
\left(\begin{array}{cc}
\alpha_r^* & \beta_r^* \\
-\beta_r & \alpha_r
\end{array}\right)
\left(\begin{array}{cc}
\omega_k &  0 \\
0 & -\omega_k
\end{array}\right)
\left(\begin{array}{cc}
\alpha_r & -\beta_r^* \\
\beta_r & \alpha^*_r
\end{array}\right)\,,
\end{eqnarray}
that is, 
\begin{align}\label{eq:alphabeta}
|\alpha_r|^2-|\beta_r|^2={\cal A}_r/\omega_k\,, \qquad 2\,\alpha_r\,\beta_r=-{\cal B}_r/\omega_k\,.
\end{align}

This way, if one defines new creation/annihilation operators 
\begin{eqnarray}
\left(\begin{array}{c}
\hat{A}_r(\bk,\,\tau) \\ 
\hat{B}_r^\dagger(-\bk,\,\tau)
\end{array}\right)\equiv
\left(\begin{array}{cc}
\alpha_r(k,\,\tau) & -\beta_r^*(k,\,\tau) \\
\beta_r(k,\,\tau) & \alpha^*_r(k,\,\tau)
\end{array}\right)
\left(\begin{array}{c}
\hat{a}_r(\bk) \\ 
\hat{b}_r^\dagger(-\bk)
\end{array}\right)\,,
\end{eqnarray}
the Hamiltonian takes the form
\begin{align}
\hat{H} = \int d^3\bk\,\omega_k(\tau)\left[\hat{A}_r^\dagger(\bk,\,\tau)\hat{A}_r(\bk,\,\,\tau)-\hat{B}_r(\bk,\,\tau)\hat{B}_r^\dagger(\bk,\,\tau)\right]\,,
\end{align}
so that we can interpret $\hat{A}^\dagger_r(\bk,\,\tau)$ and $\hat{B}^\dagger_r(\bk,\,\tau)$ as creation operators for physical states with energy $\omega_k$ at time $\tau$. The number density of fermions with helicity $r$ and momentum $k$ is then given by
\begin{align}
\frac{\langle 0|\hat{A}_r^\dagger(\bk,\,\tau)\hat{A}_r(\bk,\,\tau)|0\rangle}{{\cal V}}=\left|\beta_r(k,\,\tau)\right|^2\,,
\end{align}
where ${\cal V}$ is the volume of space.

The normalization condition $|\alpha_r(k,\,\tau)|^2+|\beta_r(k,\,\tau)|^2=1$ implies that both the $\hat{a}_r$, $\hat{b}_r$ and the $\hat{A}_r$, $\hat{B}_r$ operators satisfy canonical commutation relations. As a consequence, one obtains the expression for the occupation number,
\begin{align}
|\beta_r(k,\,\tau)|^2=\frac12-\frac{{\cal A}_r(k,\,\tau)}{2\,\omega_k}\,.
\end{align}

Using eqs.~(\ref{eq:alphabeta}) along with the normalization condition $|\alpha_r|^2 + |\beta_r|^2 = 1$, we obtain
%
%
\begin{align}
	\begin{split}
		\alpha_r &= \frac{r}{2} \sqrt{1 - r\frac{k}{\omega_k}}\,e^{-i\theta}u_r +\frac12  \sqrt{1 + r\frac{k}{\omega_k}}\,e^{i\theta}\,v_r\,, \\
	   \beta_r &= -\frac{r}{2} \sqrt{1 + r\frac{k}{\omega_k}}e^{-i\theta}u_r + \frac12\sqrt{1 - r\frac{k}{\omega_k}}\,e^{i\theta}v_r\,.
	\end{split}
\end{align}

Inserting the above expressions into the equations of motion~(\ref{eq:eom_uv}) we obtain the equations controlling the evolution of the Bogolyubov coefficients
\begin{align}\label{eq:eom_ab}
\alpha_r'&=-i\left(\omega-k\,r\frac{{\theta'}}{\omega}\right)\alpha_r + m\left(-k\,r\frac{{a'}}{2\,\omega^2}+i\,a\,\frac{{\theta'}}{\omega}\right)\beta_r\nonumber\\
\beta_r'&= m\left(k\,r\frac{{a'}}{2\,\omega^2}+i\,a\frac{{\theta'}}{\omega} \right)\alpha_r+i\left(\omega-k\,r\,\frac{{\theta'}}{\omega}\right)\beta_r\,.
\end{align}

Eqs.~(\ref{eq:eom_ab}) show that no particle production occurs in the limit $m\to 0$, as the coefficient of $\alpha_r$ in the equation for $\beta_r'$ vanishes in this limit. This is consistent with the fact that in this limit all coupling to $\theta_m$ and $\theta_5$ can be eliminated with a chiral transformation. 

\section{Calculations in radiation/matter dominated Universe}
\label{sec:solutions}%

In this section, we provide solutions to the master equations~(\ref{eq:eom_ab}) for the Bogolyubov coefficients to derive the total number of fermions produced in cosmological regimes - radiation and matter domination - of physical interest. 

\subsection{Radiation domination}

To start with, we assume that the Universe is radiation-dominated, denoting with a subscript ${}_{\rm {RD,\,end}}$ all quantities evaluated at the end of this period. In particular, $H_{\rm {RD,\,end}}$ denotes the value of the Hubble parameter at the end of the radiation-dominated regime, and we set $a_{\rm {RD,\,end}}=1$.

The scale factor is given by
\begin{equation}
a(\tau) = \frac{\tau}{\tau_{\rm {RD,\,end}}} = H_{\rm {RD,\,end}}\,\tau = (2\,H_{\rm {RD,\,end}}\,t)^{1/2}\,,
\end{equation}
where $t$ denotes the cosmic time, $dt=a(\tau)\,d\tau$. The equations for the Bogolyubov coefficients can be written using the scale factor $a$ as an independent variable and read 
\begin{align}\label{eq:eom_RD}
\frac{d\alpha_r}{da} &= -i\left(\Tilde{\omega}-r\frac{\Tilde{k}}{\tilde\omega}\frac{d\theta}{da}\right)\alpha_r +\Tilde{m}\left(-r\frac{\Tilde{k}}{2\,\Tilde{\omega}^2}+i\frac{a}{\tilde\omega}\frac{d\theta}{da}\right)\beta_r\,,\nonumber\\
\frac{d\beta_r}{da}&= \tilde{m}\left(r\frac{\tilde{k}}{2\,\tilde\omega^2}+i\,\frac{a}{\tilde\omega}\frac{d\theta}{da} \right)\alpha_r+i\left(\tilde\omega-r\frac{\tilde{k}}{\tilde\omega}\frac{d\theta}{da}\right)\beta_r\,.
\end{align}
where all the tilded parameters correspond to dimensionful quantities measured in  units of $H_{\rm {RD,\,end}}$: $\Tilde{m}\equiv m/H_{\rm {RD,\,end}}$, $\Tilde{k}\equiv k/H_{\rm {RD,\,end}}$, $\Tilde{\omega}\equiv \omega/H_{\rm {RD,\,end}}$.

Pauli blocking implies that $|\beta_r|\le 1$, and in important regions of the parameter space, the stronger condition $|\beta_r|\ll 1$ is satisfied, in which case one can neglect the terms proportional to $\beta_r$ in the right hand sides of eqs.~(\ref{eq:eom_RD}), so that the solution can be written as a single integral,
\begin{align}\label{eq:betaint_rd}
    \beta_r(a)\simeq \tilde{m}\int_0^a\,da'\left(r\frac{\tilde{k}}{2\,\tilde\omega^2}+i\,\frac{a'}{\tilde\omega}\frac{d\theta}{da'} \right)\,e^{-i\int_0^{a'} da''\,\left(\Tilde{\omega}-r\frac{\Tilde{k}}{\tilde\omega}\frac{d\theta}{da''}\right)}\,, \qquad |\beta_r(a)|\ll 1\,.
\end{align}

For a massive axion-like field $\phi$ with mass $\mu$, which we assume to be uniform in space, the equation of motion in cosmic time $t$ during radiation domination ($H=(2\,t)^{-1}$) reads
\begin{align}
\frac{d^2\phi}{dt^2}+\frac{3}{2\,t}\frac{d\phi}{dt}+\mu^2\phi=0\,,
\end{align}
whose solution, converging to $\phi_0$ as $t\to 0$, reads
\begin{align}
\phi(t)=2^{1/4}\,\Gamma(\frac54)\,\phi_0\,\frac{J_{1/4}(\mu\,t)}{(\mu\,t)^{1/4}}\Longrightarrow \theta(a)=2^{1/2}\,\Gamma(\frac54)\,\theta_0\,\frac{J_{1/4}(\tilde\mu\,a^2/2)}{\tilde\mu^{1/4}\,a^{1/2}}\,,
\end{align}
where $J_\nu(x)$ denotes the Bessel function of the first kind and where $\theta_0\equiv \theta(a\to 0)$.

Since the Hubble parameter at the end of the radiation-dominated regime is of the order of $10^{-27}$~eV, which is tiny compared to particle physics scales, we will study the parameter space of the system in the regime  $\Tilde{\mu}\gg 1$, $\Tilde{m}\gg 1$.

We can find approximate formulae for the spectrum of produced fermions at the end of the radiation-dominated period, $a=1$, by estimating the integral~(\ref{eq:betaint_rd}). In the Appendix, we show an example (on a matter-dominated background) of the calculation leading to such a spectrum. If the integral is of the order of unity or larger, we set $|\beta_r|\simeq 1$. Due to the large number of regimes of parameters to survey, we focus only on the case $\theta_0={\cal O}(1)$, even if $\theta_0\gg 1$ is also possible.

We first analyze the region of parameter space where $1\ll \Tilde{m}\ll \Tilde{\mu}$. In this regime we obtain, at the end of the radiation-dominated regime, the approximate spectrum
\begin{align}
|\beta_r(1)|^2\simeq
\left\{
\begin{array}{ll}
1 &,\,\, \tilde{k}\ll\sqrt{\tilde{m}}\\
\dfrac14\dfrac{\tilde{m}^2}{\tilde{k}^4}\left|e^{-i\tilde{k}^2/\tilde{m}}-1+e^{-ir\theta_0}\right|^2 &,\,\, \sqrt{\tilde{m}}\ll\tilde{k}\ll\sqrt{\tilde{\mu}}\,,\\
4\,\dfrac{\tilde{m}^2}{\tilde{k}\,\tilde{\mu}^{3/2}}\,\theta_0^2\,\Gamma(\frac54)^2 & ,\,\,\sqrt{\tilde{\mu}}\ll\tilde{k}\ll\tilde{\mu}\,,\\
\dfrac{8}{\pi}\,\dfrac{\tilde{m}^2\,\tilde{\mu}^{1/2}}{\tilde{k}^4}\,\theta_0^2\,\Gamma(\frac54)^2\,\cos^2\left(\dfrac{\tilde{\mu}}{2}+\dfrac{\pi}{8}\right) &,\,\, \tilde{k}\gg\tilde{\mu}\,.
\end{array}
\right.
\end{align}

By inspecting the dependence of the spectrum on the parameter $\theta_0$, we note that for momenta $\tilde{k}$ smaller than $\sqrt{\tilde{\mu}}$, fermions are predominantly produced by the expansion of the Universe, whereas at larger momenta the oscillating axion is the dominant source of matter production. Remembering that we are assuming $\theta_0={\cal O}(1)$, we thus see that fermions are predominantly produced by the oscillating axion.

The total number density of fermions produced at the end of radiation domination is then obtained by computing the integral $ N_{\rm {RD,\,end}}=\int d^3\bk\,|\beta_r(1)|^2$. Using the analytical expression for the spectra given above, we obtain 
\begin{align}\label{eq:ntot_rd_smallm}
&N_{\rm {RD,\,end}}\simeq 20\,m^2\,\sqrt{\mu\,H_{\rm {RD,\,end}}}\,\theta_0^2\,,\nonumber\\
&\qquad\qquad (1\ll \Tilde{m}\ll \Tilde{\mu}\,,\quad \theta_0={\cal O}(1))
\end{align}
where we have kept only the leading term in the $1\ll \Tilde{m}\ll \Tilde{\mu}$ limit. 

A similar analysis in the regime $1\ll \tilde{\mu}\ll \Tilde{m}$ gives the approximate spectrum
\begin{align}
|\beta_r(1)|^2=
\left\{
\begin{array}{ll}
1 &,\,\, \tilde{k}\ll\sqrt{\tilde{m}}\,,\\
\dfrac14\dfrac{\tilde{m}^2}{\tilde{k}^4}\left|e^{-i\tilde{k}^2/\tilde{m}}-1-e^{-i\tilde{k}^2/2\tilde{m}}\right|^2 &,\,\, \sqrt{\tilde{m}}\ll k\ll \tilde{m}/\sqrt{\tilde{\mu}}\,,\\
\dfrac{8}{\pi}\,\dfrac{\tilde{\mu}^{1/2}\,\tilde{m}}{\tilde{k}^{3}}\,\theta_0^2\,\Gamma^2(\frac54)\,\cos^2\Big(\dfrac{\tilde{\mu}\,\tilde{k}^2}{2\,\tilde{m}^2}+\dfrac{\pi}{8}\Big) &,\,\, \tilde{m}/\sqrt{\tilde{\mu}}\ll \tilde{k}\ll \tilde{m}\,,\\
\dfrac{8}{\pi}\dfrac{\tilde{\mu}^{1/2}\,\tilde{m}^2}{\tilde{k}^4}\,\theta_0^2\,\Gamma^2(\frac54)\,\cos^2\Big(\dfrac{\tilde{\mu}}{2}+\dfrac{\pi}{8}\Big) &,\,\, k\gg \tilde{m}\,.
\end{array}
\right.
\end{align}

Again, axion-induced production dominates at larger momenta ($k\gtrsim m\sqrt{H_{\rm RD,\,end}/\mu}$, in this case). Unlike what happens in the case $1\ll \Tilde{m}\ll \Tilde{\mu}$, however, the spectrum has a steep slope at those large momenta. As a consequence, the total number of produced fermions is dominated by gravitational effects. Keeping also the leading term proportional to $\theta_0^2$, the total number of fermions is given by the formula
\begin{align}\label{eq:ntot_rd_largem}
&N_{\rm {RD,\,end}}\simeq .9\,(m\,H_{\rm {RD,\,end}})^{3/2}+20\,m\,\sqrt{\mu\,H_{\rm {RD,\,end}}^3}\,\theta_0^2\,\log\left(\sqrt{m/\mu}/\theta_0^2\right)\,,\nonumber\\
&\qquad\qquad (1\ll \Tilde{\mu}\ll \Tilde{m}\,,\quad \theta_0={\cal O}(1))\,.
\end{align}

The behavior $N_{\rm {RD,\,end}}\propto m^{3/2}$ for the component of fermion production induced by the expansion of the Universe can also be easily derived as follows. Dimensional analysis, and the fact that fermions coupled to gravity only become conformal (and therefore insensitive to the expansion of the Universe) in the massless limit, implies that fermions are produced by the expanding Universe when the Hubble parameter is of the order of $m$.  Since the only relevant dimensionful quantity is $m$, the number density of the fermions at the time of their production must scale as  $N(H=m) \simeq m^3$. At later times ($H<m$), the fermions will just be diluted by the expansion of the Universe:
\begin{equation}
	N(H<m) \simeq m^3 \left(\frac{a(H=m)}{a(H)}\right)^3\,.
\end{equation}
For radiation domination $H\propto 1/a^2$, so that $a(H)\propto H^{-1/2}$. We thus obtain the result $N\propto m^{3/2}\,H_{\rm {RD,\,end}}^{3/2}$.

\begin{figure}
\centering
\includegraphics[width=.45\textwidth]{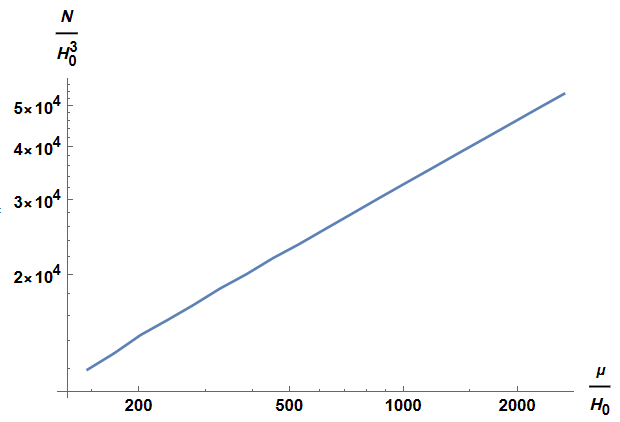}
\hspace{.5cm}
\includegraphics[width=.45\textwidth]{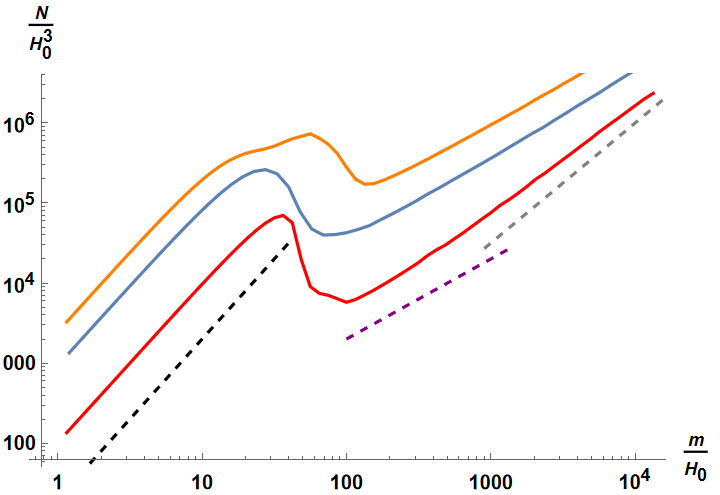}
\caption{The total number density $N_{\rm {RD,\,end}}$ of produced fermions with helicity $r=+1$ during radiation domination. Left panel: the dependence of $N/H_{\rm {RD,\,end}}^3$ on $\mu/H_{\rm {RD,\,end}}$ for $m=10\,H_{\rm {RD,\,end}}$ and $\theta_0=1$. Right panel: the dependence of $N/H_{\rm {RD,\,end}}^3$ on $m/H_{\rm {RD,\,end}}$ for $\mu=100\,H_{\rm {RD,\,end}}$ and (solid lines, bottom to top) for $\theta_0=1,\,3,\,5$. The dashed lines, left to right, show the powers $\tilde{m}^2$,    $\tilde{m}^1$, $\tilde{m}^{3/2}$.}
\label{fig:f010203}
\end{figure}

We have verified numerically the validity of the formulae given above for the occupation number which is illustrated in Figure~\ref{fig:f010203} In the left panel, we show a log-log plot of the number of produced particles when $a=1$ for $m=10\,H_{\rm {RD,\,end}}$, $\theta_0 = 1$, and varying $\mu$, $150\,H_{\rm {RD,\,end}}\le \mu\le 2500\,H_{\rm {RD,\,end}}$. We have checked that the slope of the line is $1/2$ and remains constant in the range $\Tilde{m}\ll\Tilde{\mu}$ at least in the regime $1<\theta_0<5$.

The right panel shows the log-log plot of the  number density of produced fermions, as a function of the mass $m$ and for fixed $\mu=100\,H_{\rm {RD,\,end}}$, for values of $\theta_0=1,\,3,\,5$. In particular, this plot shows that in the region $\Tilde{m}\ll \Tilde{\mu}$ the number density of produced fermions is proportional to $\tilde{m}^2\,\theta_0^2$. We also find an intermediate regime with moderately large $\tilde{m}\gtrsim \tilde{\mu}$  where we reobtain the behavior $N_{\rm {RD,\,end}}\propto \tilde{m}\,\theta_0^2$ shown in eq.~(\ref{eq:ntot_rd_largem}). In this regime fermion production is still dominated by the oscillations of the pseudoscalar (as opposed to be due  to the expansion of the Universe), and we see in fact that the number of produced fermions is proportional to $\theta_0^2$. As a consequence, as we increase $\theta_0$, the $N_{\rm {RD,\,end}}\propto \tilde{m}^{3/2}$ component (which is also present in eq.~(\ref{eq:ntot_rd_largem})) starts dominating at increasingly higher values of $\Tilde{m}$. For instance, one cannot see the $N_{\rm {RD,\,end}}\propto \tilde{m}^{3/2}$ behavior in Figure~\ref{fig:f010203}  for $\theta_0 =3,\, 5$, since this behavior kicks in at values of $\tilde{m}$ that are too large to be covered by our numerical evaluations.

Finally, we note that the left- and right-handed fermions will be produced at a similar rate. This is because the handedness of the produced fermions depends on the sign of $d\theta/da$, and, since its mass is very large, the axion oscillates at a fast rate, effectively canceling any helicity dependence.

\subsection{Matter domination}
\label{subsec:md}

For simplicity, in this subsection, we ignore the existence of a dark energy-dominated epoch and we assume that the Universe is matter-dominated from the time of matter-radiation equality until the present time. Then, denoting by $H_0$ the present value of the Hubble parameter, the scale factor is given by
\begin{align}
a(\tau)=\left(\frac{H_0\,\tau}{2}\right)^2\Longrightarrow a(t)=\left(\frac32\,H_0\,t\right)^{2/3}\,,
\end{align}
where we have set $a=1$ at present. Also for a matter-dominated Universe, it is convenient to write the equations of motion~(\ref{eq:eom_ab}) for the Bogolyubov coefficients using the scale factor $a$ as the independent variable, which then read
\begin{align}\label{eq:bogos_md}
		\frac{d\alpha_r}{da}&=-i\left(\frac{\tilde\omega}{\sqrt{a}}-r\frac{\tilde{k}}{\tilde\omega}\frac{d\theta}{da}\right)\alpha_r + \tilde{m}\left(-r\frac{\tilde{k}}{2\,\tilde\omega^2}+i\frac{a}{\tilde\omega}\,\frac{d\theta}{da}\right)\beta_r\,,\nonumber\\
		\frac{d\beta_r}{da}&= \tilde{m}\left(r\frac{\tilde{k}}{2\,\tilde\omega^2}+i\frac{a}{\tilde\omega}\,\frac{d\theta}{da} \right)\alpha_r+i\left(\frac{\tilde\omega}{\sqrt{a}}-r\frac{\tilde{k}}{\tilde\omega}\frac{d\theta}{da}\right)\beta_r\,,
\end{align}
where all tilded quantities are understood to be the corresponding dimensionful quantities measured in units of $H_0$.

Similar to radiation domination, approximate solutions to~(\ref{eq:eom_ab}) can be obtained in the regime $|\beta_r|\ll 1$. In this case we can ignore the terms proportional to $\beta_r$ on the right hand sides of eqs.~(\ref{eq:bogos_md}), and we obtain
\begin{align}\label{eq:betaint_md}
\beta_r(a)=\tilde{m}\int_0^a\,da'\left(r\frac{\tilde{k}}{2\,\tilde\omega^2}+i\,\frac{a'}{\tilde\omega}\frac{d\theta}{da'} \right)\,e^{-i\int_0^{a'} da''\left(\frac{\Tilde{\omega}}{\sqrt{a''}}-r\frac{\Tilde{k}}{\tilde\omega}\frac{d\theta}{da''}\right)}\,, \qquad |\beta_r(a)|\ll 1\,.
\end{align}

For a massive axion-like field $\phi$ with mass $\mu$, assuming that $\phi$ is constant in space, the equation of motion in cosmic time $t$ during matter domination, $H=2/(3\,t)$, reads
\begin{align}
\frac{d^2\phi}{dt^2}+\frac{2}{t}\frac{d\phi}{dt}+\mu^2\phi^2=0\,,
\end{align}
whose solution, regular as $t\to 0$, reads
\begin{align}
\phi(t)=\phi_0\,\frac{\sin(\mu\,t)}{\mu\,t}\Longrightarrow \theta(a)=\theta_0\,\frac{\sin\left(\frac23\,\tilde\mu\,a^{3/2}\right)}{\frac23\,\tilde\mu\, a^{3/2}}\,.
\end{align}

The system thus depends on the dimensionless parameters $\tilde{m}$, $\tilde\mu$ and $\theta_0$. Given the smallness of $H_0$, the natural values for $\tilde{m}$ and $\tilde{\mu}$ are many orders of magnitude larger than unity. However, for $\tilde{\mu}\gg 1$, most of the oscillations of the axion-like field will have taken place during the radiation-dominated period. Only the relatively narrow window $1\ll \tilde\mu\lesssim 10^3$ corresponds to a regime in which oscillations, and consequently fermion production, take place during matter domination. Given the smallness (on a log scale) of the size of this window, we will neglect this possibility. We will focus instead on the phenomenologically interesting situation in which the axionic field is slowly rolling today, considering the case in which the axion is acting as dark energy. As we noted above, we assume that the Universe is matter dominated all along, so this corresponds to the regime in which the axion-like dark energy did not come to dominate the energy in the Universe yet.

Since the axion is slowly rolling, we can approximate $\mu t\ll 1$, leading to $\phi(t)\simeq \phi_0\,\left(1-\frac16 \,\mu^2\,t^2\right)$, and the equation of state parameter
\begin{align}
w(t)=\frac{\dot\phi^2-\mu^2\phi^2}{\dot\phi^2+\mu^2\phi^2}\simeq -1+\frac49\mu^2\,t^2\,,
\end{align}
which gives a current (i.e., at $t=t_0=\frac{2}{3H_0}$) value of $w_0=-1+\frac{16}{81}\tilde\mu^2$. To fix ideas, we set $\tilde\mu=.7$, which gives an equation of state parameter $w_0\simeq -.9$. In the following analytical calculations, we will keep only the leading order terms in $\tilde{\mu}$. In particular, we will set
\begin{align}
\theta(a)= \theta_0\left(1-\frac{2}{27}\tilde{\mu}^2\,a^3\right)\,.
\end{align}

Unlike the radiation-dominated case, in this subsection, we will allow the quantity $\theta_0$ to be either of the order of unity or much larger than one. In fact, while the simplest expectation for the axion-like dark energy field is a cosine potential $V(\phi)\propto \cos(\phi/f)$, in which case $\theta_0$ is typically expected to be smaller than unity, in models with monodromy the axion potential can extend on a range $\Delta\phi\gg f$ with a quadratic~\cite{Kaloper:2008qs} or more complicated~\cite{DAmico:2018mnx} potentials.

The integral~(\ref{eq:betaint_md}) can be evaluated analytically to give the spectrum of produced particles at the end of the matter domination period ($a = 1$). We will set $|\beta_r|\simeq 1$ if the integral is order of or larger than unity. 

First, we focus on the region in parameter space where $1\lesssim\Tilde{\mu}^2\,\theta_0\ll\Tilde{m}$. The approximate spectrum in this region is found to be
\begin{align}
|\beta_r(1)|^2=
\left\{
\begin{array}{ll}
1\,, &,\,\, \Tilde{k}\ll \Tilde{m}^{1/3}\\ 
\vspace{0.1cm}
\dfrac14\dfrac{\Tilde{m}}{\Tilde{k}^3} & ,\,\,\Tilde{m}^{1/3} \ll \Tilde{k} \ll \dfrac{\Tilde{m}}{\Tilde{\mu}^{4/3}\,\theta_0^{2/3}}\ll \Tilde{m},\\
\vspace{0.1cm}
\dfrac{4}{81}\,\dfrac{\Tilde{\mu}^4\,\theta_0^2}{\Tilde{m}^2} &,\,\, \Tilde{m}^{1/3} \ll \dfrac{\Tilde{m}}{\Tilde{\mu}^{4/3}\,\theta_0^{2/3}}\ll \Tilde{k} \ll \Tilde{m},\\
\vspace{0.1cm}
\dfrac{4}{81}\,\dfrac{\Tilde{m}^2\,\Tilde{\mu}^4}{\Tilde{k}^4}\,\theta_0^2 &,\,\, \Tilde{m} \ll \Tilde{k}\,.
\end{array}
\right.
\end{align}

These expressions show that for momenta smaller than $m\,H^{4/3}_0/(\mu^{4/3}\,\theta_0^{2/3})$ the fermions are predominantly produced by gravitational effects, whereas in the regime of larger momentum the slow rolling axions enhances the production of fermions. 

Our analytical expressions for the spectra can be used to give the total number density of fermions produced by computing $N_0=\int d^3\bk\,|\beta_r(1)|^2$. To leading order, the number density reads
\begin{align}\label{eq:finnum_md_mlarge}
&N_0\simeq .2\,\frac{m\,\mu^4}{H_0^2}\,\theta_0^2\,,\nonumber\\
&\qquad\qquad (1\lesssim \tilde{\mu}^2\,\theta_0 \ll \Tilde{m})\,.
\end{align}

In the regime $1 \ll\Tilde{m} \ll \tilde{\mu}^2\,\theta_0$, instead, we obtain the following approximate spectra:
\begin{align}
\label{eq:specmd41}
|\beta_r(1)|^2\simeq 
\left\{
\begin{array}{ll}
1 &,\,\, \Tilde{k}\ll\Tilde{m}\\ 
\vspace{0.1cm}
\dfrac{\Tilde{m}^2}{\Tilde{k}^2}\dfrac{1+r}{2}+{\rm Max}\left\{\dfrac{18\,\pi}{5}\,\dfrac{\tilde{m}^2}{\tilde{\mu}^2\,\theta_0},\,1\right\}\times\dfrac{1-r}2 &,\,\, \Tilde{m} \ll \Tilde{k} \ll \Tilde{\mu}^2\,\theta_0,\\
\vspace{0.1cm}
\dfrac{4}{81}\,\dfrac{\Tilde{m}^2\,\Tilde{\mu}^4}{\Tilde{k}^4}\,\theta_0^2 & ,\,\,\Tilde{k}\gg\Tilde{\mu}^2\,\theta_0 \, .\\
\end{array}
\right.
\end{align}

The fact that $\theta'$ has a definite sign leads to a breaking of parity, so that the number of left-handed and of right-handed fermions will be different. In fact, the spectrum has different scaling depending on whether $r=1$ or $r=-1$. To leading order the total number density reads
\begin{align}\label{eq:finnum_md_msmall}
&N_{\rm {MD,\,end}}\simeq \left\{
\begin{array}{ll}
5.\,\dfrac{m^2\,\mu^2}{H_0}\,\theta_0\,\dfrac{1+r}{2} + 50.\times\dfrac{m^2\,\mu^4}{H_0^3}\,\theta_0^2\,\dfrac{1-r}{2} &,\,\, 1\ll \Tilde{m}\ll \Tilde{m}^2 \ll \tilde{\mu}^2\,\theta_0\,,\\
\vspace{.2cm}
5.\,\dfrac{m^2\,\mu^2}{H_0}\,\theta_0\,\frac{1+r}{2} + .3\,\dfrac{\mu^6}{H_0^3}\,\theta_0^3\,\dfrac{1-r}{2} &,\,\, 1\ll \Tilde{m}\ll \tilde{\mu}^2\,\theta_0\ll \Tilde{m}^2\,,
\end{array}
\right.
\end{align}

\begin{figure}[h]
\centering
\includegraphics[width=.4\textwidth]{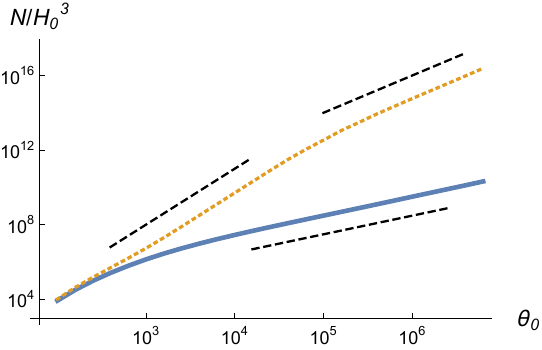}
\hspace{1cm}
\includegraphics[width=.4\textwidth]{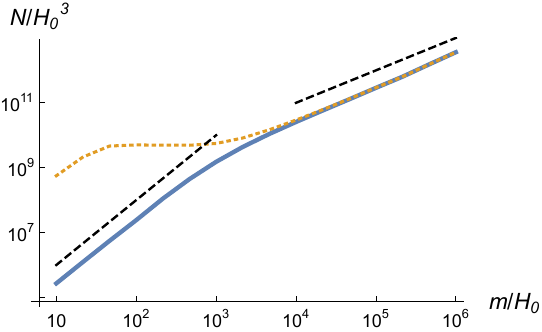}
\caption{The final number density of $r=-1$ (top, short dashed curves) and $r=+1$ (bottom, solid curves) produced fermions, in the case of a slowly-rolling axion-like field with mass $\mu=.7\,H_0$. Left panel: the dependence on $\theta_0$ for fixed $m=100\,H_0$. The long dashed lines indicate the scalings $\propto \theta_0^3$ (left), $\propto \theta_0^2$ (top right) and $\propto \theta_0$ (bottom right). Right panel: the dependence on $m/H_0$ for fixed $\theta_0=10^4$. The long dashed lines in the right panel show the powers $\tilde{m}^2$ (left), and $\tilde{m}^1$ (right).}
\label{fig:mdmup7m103104}
\end{figure}

We have checked numerically the validity of the scalings in eqs.~(\ref{eq:finnum_md_mlarge}) and~(\ref{eq:finnum_md_msmall}). The left panel of Figure~\ref{fig:mdmup7m103104} shows that the number density of the fermions with helicity that are produced most efficiently scales as $\theta_0^3$ for $\tilde{m}\lesssim\tilde{\mu}^2\,\theta_0\lesssim \tilde{m}^2$ and as $\theta_0^2$ for $\tilde{m}\lesssim\tilde{m}^2\lesssim \tilde{\mu}^2\,\theta_0$, whereas the number density of the fermions of the other helicity is subdominant and proportional to $\theta_0$, which is consistent with the behavior found in eq.~(\ref{eq:finnum_md_msmall}). The right panel shows a log-log plot of the total number density of fermions, as a function for $m/H_0$, obtained at $a=1$ for $\theta_0=10^4$. Consistently with eq.~(\ref{eq:finnum_md_msmall}), the total number of produced fermions of both helicities is proportional to $\tilde{m}^2$ in the regime $\tilde{m}\ll \tilde{\mu}^2\,\theta_0$ and, consistently with eq.~(\ref{eq:finnum_md_mlarge}), it is proportional to $\tilde{m}$ for $\tilde{m}\gg \tilde{\mu}^2\,\theta_0$.

\section{Applications and summary}
\label{sec:summary}%

In this work we have presented, for the first time, the equations controlling the generation of fermions due to their coupling to a time-dependent axion, eqs.~(\ref{eq:eom_ab}). From these equations, we have found approximate analytical expressions for the spectra of the produced fermions in the case of an axion performing many oscillations in a radiation-dominated Universe and for a slowly rolling axion in a matter-dominated Universe. Here we discuss some possible applications and extensions of our work.

A possible extension of this setup concerns the possibility of modifying the cosmological constraint on $f=f_{\rm QCD\ axion}$ in the case where the axion is the QCD axion. In this case, if the standard misalignment mechanism (where $\phi$ starts oscillating with an amplitude of the order of $f_{\rm QCD\ axion}$) is realized, then the oscillating axion will overclose the Universe if $f_{\rm QCD\ axion}$ is larger than $10^{12}$~GeV or so. This constraint, about $4$ orders of magnitude tighter than the ``natural'' value obtained for the axion decay constants in UV-complete theories~\cite{Svrcek:2006yi}, might then be modified if fermion production drains energy from the oscillating axion. A similar idea has been explored, in the case in which vectors are produced by the rolling axion, in~\cite{Agrawal:2017eqm,Kitajima:2017peg}. The produced fermions, which are not necessarily Standard Model particles, might then decay into light Standard Model degrees of freedom that will thermalize, effectively disappearing from our spectrum. This analysis requires the study of the evolution of the axionic condensate in the strong backreaction regime, which we have not considered in the present paper.

Moving to the more recent, matter-dominated Universe, one might wonder whether in the scenario considered in Section~\ref{subsec:md} the produced fermions might be identified with Standard Model neutrinos. In this case, neutrino production would occur efficiently well after Big Bang Nucleosynthesis and the decoupling era. More in general, one might wonder if axionic quintessence might generate a sizable amount of nonrelativistic fermions in the late Universe, similarly to the scenario discussed in~\cite{Berghaus:2019cls}. From eq.~(\ref{eq:finnum_md_mlarge}), and since the fermions produced this way are nonrelativistic, we can estimate the total energy density of relatively heavy fermions to be of the order of $m^2\,\mu^4\,\theta_0^2/H_0^2$, giving a contribution to $\Omega\approx (m/M_P)^2\,(\mu/H_0)^4\,\theta_0^2$ with $\mu/H_0={\cal O}(1)$. In models of axion quintessence with monodromy, the quantity $\theta_0$ can be as large as $M_P/f$. We thus get that the contribution to $\Omega$ from the produced fermions could be of the order of $m^2/f^2$, which can be sizable if $m$ is close to $f$.

\acknowledgments We thank Yohei Ema and Valerie Domcke for very interesting discussions.  This work is partially supported by the US-NSF grant PHY-2112800.

\appendix

\section{Derivation of the fermion spectrum for $1 \ll\Tilde{m} \ll \tilde{\mu}^2\,\theta_0$ in matter-domination}
\label{App:A}

In this appendix we show the main steps leading to the spectrum of particles in matter domination given by eq.~(\ref{eq:specmd41}), limiting ourselves to the regime  $\Tilde{m} \ll \tilde{\mu}^2\,\theta_0$.  The remaining spectra presented in the main body of the paper are obtained using similar techniques. 

It turns out that to compute the integral~(\ref{eq:betaint_md}) it is necessary to distinguish the regimes $\Tilde{m} \ll \tilde{\mu}^2\,\theta_0 \ll \tilde{m}^2$ and $\Tilde{m} \ll \tilde{m}^2 \ll \tilde{\mu}^2\,\theta_0$ (remember that $\tilde{m}\gg 1$). However, we find that the spectra look the same in both regimes, except for one difference that we will highlight below.  For this reason, here we will show the derivation when $\Tilde{m} \ll \tilde{\mu}^2\,\theta_0 \ll \tilde{m}^2$. 

Using
\begin{align}
\frac{d\theta}{da}= -\frac{2}{9}\,\tilde{\mu}^2\,\theta_0\,a^2\,,
\end{align}
the integrand~(\ref{eq:betaint_md}) can be approximated as
\begin{align}
&\tilde{m}\Big(r\frac{\tilde{k}}{2\,\tilde\omega^2}+i\,\frac{a'}{\tilde\omega}\frac{d\theta}{da'} \Big)e^{-i\int_0^{a'} da''\big(\frac{\Tilde{\omega}}{\sqrt{a''}}-r\frac{\Tilde{k}}{\tilde\omega}\frac{d\theta}{da''}\big)}\nonumber\\
&\qquad \simeq 
\left\{
\begin{array}{ll}
\dfrac{\tilde{m}}{\tilde{k}}\left(\dfrac{r}{2}-i\,\Theta_0\,a'{}^3 \right)\,e^{-i(2\tilde{k}\sqrt{a'}+\frac{r}{3}\,\Theta_0\,a'{}^3)} &,\,\, 0<a'\ll \tilde{k}/\tilde{m}\,,\\
\,\left(\dfrac{\tilde{r\,k}}{2\,\tilde{m}\,a'{}^2}-i\,\Theta_0\,a'{}^2 \right)\,e^{-i(\frac23\,\tilde{m}\,a'{}^{3/2}+\frac{r}{2}\,\Theta_0\,\frac{\tilde{k}}{\tilde{m}} a'{}^2)} &,\,\, \tilde{k}/\tilde{m}\ll a'<1\,,
\end{array}\right.
\end{align}
and we have defined 
\begin{align}
\Theta_0\equiv \frac{2}{9}\tilde{\mu}^2\,\theta_0\,,
\end{align}
and where the second approximation can be realized only for $\tilde{k}\ll \tilde{m}$.

Next we observe that in the case $a'\ll \tilde{k}/\tilde{m}$ the first term in the exponent dominates over the second for $a'\ll (k/\Theta_0)^{2/5}$, whereas in the case $\tilde{k}/\tilde{m}\ll a'<1$ the first term in the exponent dominates over the second for $a'\ll \tilde{m}^4/(\tilde{k}^2\,\Theta_0^2)$. These two values of the scale factor cross $a'=\tilde{k}/\tilde{m}$ when $\tilde{k}\simeq \tilde{m}^{5/3}/\Theta_0^{2/3}$. Remembering that we are assuming $\Tilde{m} \ll \Theta_0 \ll \tilde{m}^2$, we have the hierarchy
\begin{align}
    \tilde{m}^{1/3}\ll \frac{\tilde{m}^{5/3}}{\Theta_0^{2/3}}\ll \tilde{m}\ll \Theta_0 \ll \tilde{m}^2 
\end{align}

For values of $k$ within the intervals determined by the inequalities above the functions appearing in the integral can be simplified as follows:

\paragraph{1: $\tilde{k}\ll \tilde{m}^{1/3}\ll \frac{\tilde{m}^{5/3}}{\Theta_0^{2/3}}\ll \tilde{m}\ll \Theta_0 \ll \tilde{m}^2 $.} The integral can be written as:
\begin{align}\label{eq:approx_smallk}
    \beta_r(1) &\simeq \frac{\tilde{m}}{\tilde{k}}\int^{\tilde{k}/\tilde{m}}_0\,da\,\left(\frac{r}{2}-i\,\Theta_0\,a{}^3 \right)\,e^{-2i\tilde{k}\sqrt{a}}+\int_{\tilde{k}/\tilde{m}}^{\tilde{m}^4/\tilde{k}^2\theta_0^2}\,da\,\left(\frac{r\,\tilde{k}}{2\,\tilde{m}\,a^2} - i\,\Theta_0\,a^2\right)e^{-\frac{2}{3}i\tilde{m}\,a^{3/2}}\nonumber\\
    &+\int_{\tilde{m}^4/\tilde{k}^2\Theta_0^2}^1\,da\,\left(\frac{r\,\tilde{k}}{2\,\tilde{m}\,a^2} - i\,\Theta_0\,a^2\right)e^{-ir\frac{\tilde{k}\,\Theta_0\,a^2}{2\,\tilde{m}}}\,,
\end{align}
where the exponent in the first integral is  smaller than $2\,\tilde{k}^{3/2}/\tilde{m}^{1/2}$ which is much smaller than unity. As a consequence the first term in the first integral evaluates to 
\begin{align}
\frac{\tilde{m}}{\tilde{k}}\int^{\tilde{k}/\tilde{m}}_0\,da\,\frac{r}{2}\,e^{-2i\tilde{k}\sqrt{a}}\simeq \frac{\tilde{m}}{\tilde{k}}\int^{\tilde{k}/\tilde{m}}_0\,da\,\frac{r}{2}=\frac{r}{2}\,,
\end{align}
which is ${\cal O}(1)$. Since we do not expect the other terms to cancel the first one (particles are generally not reabsorbed by the axion condensate after being created), we can stop evaluating our integral here and declare that in this region of parameter space $|\beta_r|={\cal O}(1)$.

\paragraph{2: $\tilde{m}^{1/3}\ll \tilde{k}\ll \frac{\tilde{m}^{5/3}}{\Theta_0^{2/3}}\ll \tilde{m}\ll \Theta_0 \ll \tilde{m}^2 $.} The integral takes the same form as in eq.~(\ref{eq:approx_smallk}) above, but we cannot assume the first phase to be negligible anymore. However, by substituting $\frac{2}{3}\,\tilde{m}\,a^{3/2} \equiv x$ in the second term in the second integral, we get
\begin{align}
    &- i\,\Theta_0\,\int_{\tilde{k}/\tilde{m}}^{\tilde{m}^4/\tilde{k}^2\theta_0^2}\,da\,a^2\,e^{-\frac{2}{3}i\tilde{m}\,a^{3/2}} =-\frac32 i\frac{\Theta_0}{\tilde{m}^2}\int_{\frac23\tilde{k}^{3/2}/\tilde{m}^{1/2}}^{\frac23\tilde{m}^7/(\tilde{k}\,\Theta_0)^3}\,dx\,x\,e^{-ix} \simeq \frac{\tilde{m}^5}{\tilde{k}^3\,\Theta_0^2}\,e^{-i\frac23\frac{\tilde{m}^7}{\tilde{k}^3\theta_0^3}}\,,
\end{align}
where in the last step we have used the fact that the integral is dominated by the upper limit which, since $\tilde{m}^2\gg\Theta_0$, is much larger than unity. In this region of parameter space, then, this contribution of the integral to $|\beta_r|$ is much greater than unity. Occupation numbers much larger than unity are forbidden by Pauli blocking, so this result shows that our approximation $|\beta_r|\ll 1$ is violated in this regime. We will therefore set $|\beta_r|={\cal O}(1)$ also in this portion of the parameter space.

\paragraph{3: $\tilde{m}^{1/3}\ll \frac{\tilde{m}^{5/3}}{\Theta_0^{2/3}}\ll \tilde{k}\ll \tilde{m}\ll \Theta_0 \ll \tilde{m}^2 $.} In this case the integral takes the approximate form
\begin{align}
&\beta_r(1)\simeq \frac{\tilde{m}}{\tilde{k}}\int^{(\tilde{k}/\Theta_0)^{2/5}}_0\left(\frac{r}{2}-i\,\Theta_0\,a{}^3 \right)\,e^{-2i\tilde{k}\sqrt{a}}+\frac{\tilde{m}}{\tilde{k}}\int_{(\tilde{k}/\Theta_0)^{2/5}}^{\tilde{k}/\tilde{m}}\left(\frac{r}{2}-i\,\Theta_0\,a{}^3 \right)\,e^{-i\frac{r}{3}\,\Theta_0\,a{}^3}\nonumber\\
&+\int^1_{\tilde{k}/\tilde{m}}\big(r\frac{ \tilde{k}}{2\,\tilde{m}\,a^2}-i\,\Theta_0\,a^2 \big)\,e^{-i\frac{r}{2}\,\Theta_0\,\frac{\tilde{k}}{\tilde{m}} a^2}\,.
\end{align}
Again, let us just consider the second part of the second integral. We can write it as
\begin{align}
-i\,\frac{\tilde{m}}{\tilde{k}}\,\Theta_0\int_{(\tilde{k}/\Theta_0)^{2/5}}^{\tilde{k}/\tilde{m}}\,a{}^3 \,e^{-i\frac{r}{3}\,\Theta_0\,a{}^3}=-i\,\frac{\tilde{m}}{\tilde{k}}\,\left(\frac{3}{\Theta_0}\right)^{1/3}\int^{\frac{\Theta_0}{3}(\tilde{k}/\tilde{m})^3}_{\frac{\Theta_0}{3}(\tilde{k}/\Theta_0)^{6/5}}dx\,x^{1/3}\,e^{-i\,r\,x}
\end{align}
that is, again dominated by its upper end, which is much larger than unity in the portion of parameter space we are exploring. We thus obtain
\begin{align}\label{eq:intcase3}
-i\,\frac{\tilde{m}}{\tilde{k}}\,\Theta_0\int_{(\tilde{k}/\Theta_0)^{2/5}}^{\tilde{k}/\tilde{m}}\,a{}^3 \,e^{-i\frac{r}{3}\,\Theta_0\,a{}^3}\simeq r\,e^{-ir\frac{\Theta_0}{3}\,\frac{k^3}{m^3}}={\cal O}(1)\,,
\end{align}
so that, also in this regime, we can set $|\beta_r|={\cal O}(1)$.

\paragraph{4: $\tilde{m}^{1/3}\ll \frac{\tilde{m}^{5/3}}{\Theta_0^{2/3}}\ll \tilde{m}\ll \tilde{k}\ll \Theta_0 \ll \tilde{m}^2 $.} Since $\tilde{k}>\tilde{m}$, we can find a unique approximate form for the integrand in the entire range $0<a<1$:
\begin{align}
    \beta_r &\simeq \frac{\tilde{m}}{\tilde{k}}\int_0^1\,da\,\left(\frac{r}{2} - i\,\Theta_0\,a^3\right)\,e^{-2i\tilde{k}\sqrt{a}-i\frac{r}{3}\Theta_0\,a^3}\,.
\end{align}

For $r=-1$ we can estimate this integral using the saddle point approximation. The phase is $\phi(a) = 2\tilde{k}\sqrt{a} - \frac{\Theta_0}{3}a^3$ with saddle point at $a_S = (\tilde{k}/\Theta_0)^{2/5}$. Since the integral runs between $0$ and $1$, the saddle point contributes to the integral only if $\tilde{k}<\Theta_0$. On the saddle, the phase reads, $\phi(a_S) = \frac{5}{3}\,\tilde{m}^{6/5}/\Theta_0^{1/5}$ and its second derivative is $\phi''(a_S) = -\frac{5}{2}\,\tilde{k}^{2/5}\,\Theta_0^{3/5}\gg 1$. We also note that $\Theta_0\,a_S^3=\tilde{k}^{6/5}/\Theta_0^{1/5}$. Since $\tilde{k}\gg \tilde{m}$, this implies that $\Theta_0\,a_S^3\gg \tilde{m}^{6/5}/\Theta_0^{1/5}\gg \Theta_0^{2/5}\gtrsim 1$. As a consequence, the term proportional to $-i\Theta_0 \,a^3$ in the integral dominates over that proportional to $r/2$.  By performing the Gaussian integral we then obtain
\begin{align}\label{eq:betam1saddle}
\frac{\tilde{m}}{\tilde{k}}\int_0^1\,da\,\left(\frac{r}{2} - i\,\Theta_0\,a^3\right)\,e^{-2i\tilde{k}\sqrt{a}-i\frac{r}{3}\Theta_0\,a^3}\Bigg|_{r=-1}\simeq -i\sqrt{\frac{4\,\pi}{5}}\frac{\tilde{m}}{\sqrt{\Theta_0}}e^{-i\,\frac{5}{3}\,\tilde{k}^{6/5}/\Theta_0^{1/5}-i\pi/4}\,.
\end{align}

Once again, this is in modulus much larger than unity, so in this region of parameter space we set again $|\beta_{-1}|={\cal O}(1)$.  This result is essentially the same also in the regime $\tilde{m}^2\ll\Theta_0$. In that case, however, the value of $|\beta_{-1}|$ is much smaller than $1$, so we maintain the expression $\beta_{-1}$ given by eq.~(\ref{eq:betam1saddle}).

For $r = +1$ the saddle would be at negative values of $a$, which are not sampled by our integral, so we cannot use the saddle point approximation. To find the leading behavior of the integral it is convenient to split it as
\begin{align}
     &\beta_{+1}(1)\simeq \frac{\tilde{m}}{\tilde{k}}\int_0^{(\tilde{k}/\Theta_0)^{2/5}}\,da\,\left(\frac{1}{2} - i\,\Theta_0\,a^3\right)\,e^{-2i\tilde{k}\sqrt{a}}+ \frac{\tilde{m}}{\tilde{k}}\int_{(\tilde{k}/\Theta_0)^{2/5}}^1\,da\,\left(\frac{1}{2} - i\,\Theta_0\,a^3\right)\,e^{-\frac{i}{3}\Theta_0\,a^3}\,,
\end{align}
The biggest contribution comes from the second part of the second integral above. This is the same integral as the one in eq.~(\ref{eq:intcase3}), only with a different upper limit, so we can evaluate it using the same technique. We thus obtain
\begin{align}
    \beta_+ &\simeq \frac{\tilde{m}}{\tilde{k}}e^{-i\Theta_0/3}\,,
\end{align}
whose absolute value is much smaller than unity. We thus see that in this regime there is a difference in the chirality of the produced fermions.

\paragraph{5: $\tilde{m}^{1/3}\ll \frac{\tilde{m}^{5/3}}{\Theta_0^{2/3}}\ll \tilde{m}\ll \Theta_0 \ll \tilde{k}$.} In this final region the integral reads approximately
\begin{align}
    \beta_{r}(1)\simeq \frac{\tilde{m}}{\tilde{k}}\int_0^1\,da\,\left(\frac{r}{2} - i\,\Theta_0\,a^3\right)\,e^{-2i\tilde{k}\sqrt{a}}\,,
\end{align}
where the largest contribution comes from the second term. We can use the familiar tricks to get
\begin{align}
    \beta_r(1)&\simeq \frac{\tilde{m}}{\tilde{k}^2}\,\Theta_0
\end{align}
in the entire regime $\tilde{k}\gg\Theta_0$.

\begin{figure}
\centering
\includegraphics[width=.5\textwidth]{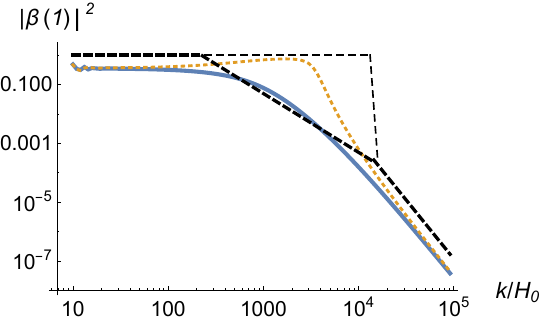}
\caption{Spectra of the $r=-1$ (top, dotted curve) and $r=+1$ (bottom, solid curve) produced fermions, in the case of a slowly-rolling axion-like field with mass $\mu=.7\,H_0$, $m=10^3\,H_0$ and $\theta_0=10^{4.5}$. The darker dashed segments correspond to the approximate expression~(\ref{eq:specmd3}) for the same choice of parameters, the thinner one for $r=-1$ and the thicker one for $r=+1$.}
\label{fig:spectrum_md_m103_f01045}
\end{figure}

To sum up, we obtain the following approximate spectra: 

\noindent {\em {(i)}} for $\tilde{m}\ll\Tilde{\mu}^2\,\theta_0 \ll\tilde{m}^2$,
\begin{align}
\label{eq:specmd3}
|\beta_r(1)|^2\simeq 
\left\{
\begin{array}{ll}
1 &,\,\, \Tilde{k}\ll\Tilde{m}\\ 
\vspace{0.1cm}
\dfrac{\Tilde{m}^2}{\Tilde{k}^2}\dfrac{1+r}{2}+\dfrac{1-r}2 &,\,\, \Tilde{m} \ll \Tilde{k} \ll \Tilde{\mu}^2\,\theta_0,\\
\vspace{0.1cm}
\dfrac{4}{81}\dfrac{\Tilde{m}^2\,\Tilde{\mu}^4}{\Tilde{k}^4}\,\theta_0^2 &,\,\, \Tilde{k}\gg\Tilde{\mu}^2\,\theta_0 \,;\\
\end{array}
\right.
\end{align}

\noindent {\em {(ii)}} for $\tilde{m}\ll\tilde{m}^2\ll\Tilde{\mu}^2\,\theta_0$, 
\begin{align}
\label{eq:specmd4}
|\beta_r(1)|^2\simeq 
\left\{
\begin{array}{ll}
1 &,\,\, \Tilde{k}\ll\Tilde{m}\\ 
\vspace{0.1cm}
\dfrac{\Tilde{m}^2}{\Tilde{k}^2}\dfrac{1+r}{2}+\dfrac{18\,\pi}{5}\,\dfrac{\tilde{m}^2}{\tilde{\mu}^2\theta_0}\dfrac{1-r}2 & ,\,\,\Tilde{m} \ll \Tilde{k} \ll \Tilde{\mu}^2\,\theta_0,\\
\dfrac{4}{81}\,\dfrac{\Tilde{m}^2\,\Tilde{\mu}^4}{\Tilde{k}^4}\,\theta_0^2 &,\,\, \Tilde{k}\gg\Tilde{\mu}^2\,\theta_0 \, .\\
\end{array}
\right.
\end{align}

We show in Figure~\ref{fig:spectrum_md_m103_f01045} the comparison between the spectra obtained by solving numerically the eqs.~(\ref{eq:bogos_md}) and the approximate expressions~(\ref{eq:specmd3}). As the figure shows, the approximate spectra match the exact ones up to ${\cal O}(1)$ factors.


\end{document}